\documentclass[%aip,
jmp,
amsmath,amssymb,
reprint,%
]{revtex4-1}
\usepackage{graphicx}% Include figure files
\usepackage{dcolumn}% Align table columns on the decimal point
\usepackage{bm}% bold math
\usepackage[utf8]{inputenc}
\usepackage[T1]{fontenc}
\usepackage{mathptmx}
\usepackage{hyperref}
\newcommand{\beq}{\begin{equation}}
\newcommand{\eeq}{\end{equation}}

\begin{document}
%\preprint{AIP/123-QED}
\title[]{Effects of oblique collimated irradiation on the onset of phototaxis-driven bioconvection in an isotropic porous medium
}
%    Force line breaks with \\	
%\author{S. K. Rajput({\dn s{\rs.} k{\rs.} rAj\7{p}t})}
\author{S. K. Rajput}
\altaffiliation[Corresponding author Electronic mail:]{ shubh.iiitj@gmail.com.}%Lines break automatically or can be forced with 
%\author{M. K. Panda({\dn m{\rs.} k{\rs.} p\2A{d}A})}
\author{M. K. Panda}%
%\email{Second.Author@institution.edu.}
\affiliation{Department of Mathematics, PDPM Indian Institute of Information Technology Design and Manufacturing, Jabalpur $482005$, India}%\\This line break forced with \textbackslash\textbackslash	
%\author{C. Author}
% \homepage{http://www.Second.institution.edu/~Charlie.Author.}
%\affiliation{%
%Second institution and/or address%\\This line break forced% with \\
%}%	
%\author{S. K. Rajput}
% \homepage{http://www.Second.institution.edu/~Charlie.Author.}
%\affiliation{Department of Mathematics, PDPM Indian Institute of Information Technology Design and Manufacturing, Jabalpur $482005$, India}
%Second institution and/or address%\\This line break forced% with \\
%}%	
%\author{}
% \homepage{http://www.Second.institution.edu/~Charlie.Author.}
%\affiliation{Department of Physics, PDPM Indian Institute of Information Technology Design and Manufacturing, Jabalpur $482005$, India}	
%\author{}
% \homepage{http://www.Second.institution.edu/~Charlie.Author.}
%\affiliation{Department of CSE, PDPM Indian Institute of Information Technology Design and Manufacturing, Jabalpur $482005$, India}	
%\date{\today}% It is always \today, today,
%but any date may be explicitly specified
%%%%%%%%%%%%%%%%%%%%%%%%%%%%%%%%%%%%%%%%%%%%%%%%%%%%%%	
\begin{abstract}
	In this study, we investigate the effects of oblique collimated irradiation on the onset of phototaxis-driven bioconvection in an isotropic porous medium. A linear stability analysis is conducted to assess the system's stability under fixed parameter values. The resulting eigenvalue problem is numerically solved using a fourth-order accurate finite difference scheme combined with Newton-Raphson-Kantorovich iterations. The results indicate that the system exhibits increased instability as the angle of incidence rises for a given Darcy number. Additionally, the critical Rayleigh number is found to be higher when a rigid top wall is considered compared to a stress-free top wall, suggesting that the suspension attains greater stability in the presence of a rigid top boundary.	
	
\end{abstract}

%%%%%%%%%%%%%%%%%%%%%%%%%%%%%%%%%%%%%%%%%%%%%%%%%%%%%%	
\maketitle
%%%%%%%%%%%%%%%%%%%%%%%%%%%%%%%%%%%%%%%%%%%%%%%%%%%%%%	
\section{Introduction}
Bioconvection is a phenomenon where self-propelled microorganisms, such as algae and bacteria, generate large-scale organized fluid motion through their collective movement~\citep{12platt1961,15bees2020,16javadi2020,13pedley1992,14hill2005, 41rajput2025d}. This flow arises due to the interplay between microbial behavior, known as taxis, and fluid dynamic instability. Taxis refers to the directed movement of microorganisms in response to environmental stimuli, including light (phototaxis), gravity (gravitaxis), fluid velocity gradients (gyrotaxis), and oxygen concentration gradients (oxytaxis). Typically, these microorganisms are slightly denser than the surrounding fluid and exhibit upward swimming behavior~\citep{13pedley1992,5vincent1996}. As they accumulate at a particular depth, they form a denser horizontal layer (hereafter referred to as the sublayer). Once the density gradient reaches a critical threshold, the sublayer destabilizes, leading to pattern formation driven by density variations within the suspension. However, pattern formation in bioconvection is not solely dependent on swimming direction or density differences. For instance, when light is incident from above, algal cells absorb and scatter light, casting shadows on the cells beneath them—a phenomenon referred to as shading. This shading effect significantly influences the morphology and dimensions of the bioconvective patterns~\citep{5vincent1996}. This study specifically investigates the role of shading in bioconvection.

To investigate bioconvective instability in a porous medium, we consider a finite-depth porous layer containing a suspension of phototactic microorganisms. This layer is illuminated from above by oblique collimated irradiation. Algal cells derive energy through photosynthesis and exhibit phototaxis, moving either toward or away from high-intensity light to maximize energy absorption or avoid photo damage, respectively. This behavior is classified as positive or negative phototaxis. As a result, the microorganisms accumulate in a sublayer at a specific depth within the suspension where the light intensity approaches a critical threshold, $G\approx G_c$, where $G_c$ represents the critical light intensity~\citep{19hader1987} [see Fig.~\ref{1}]. The position of this sublayer depends on the light intensity: it forms near the top of the layer when the intensity exceeds $G_c$ and near the bottom when the intensity is below $G_c$. This sublayer acts as a boundary between two dynamically distinct regions—a stable region above and an unstable region below. When the sublayer destabilizes, fluid motion develops in the unstable region, disrupting the stability of the overlying fluid. This process, known as penetrative convection, is commonly observed in various convective systems\citep{3ghorai2005,4panda2016}.

\begin{figure}[!htbp]
	\centering
	\includegraphics[scale=0.53]{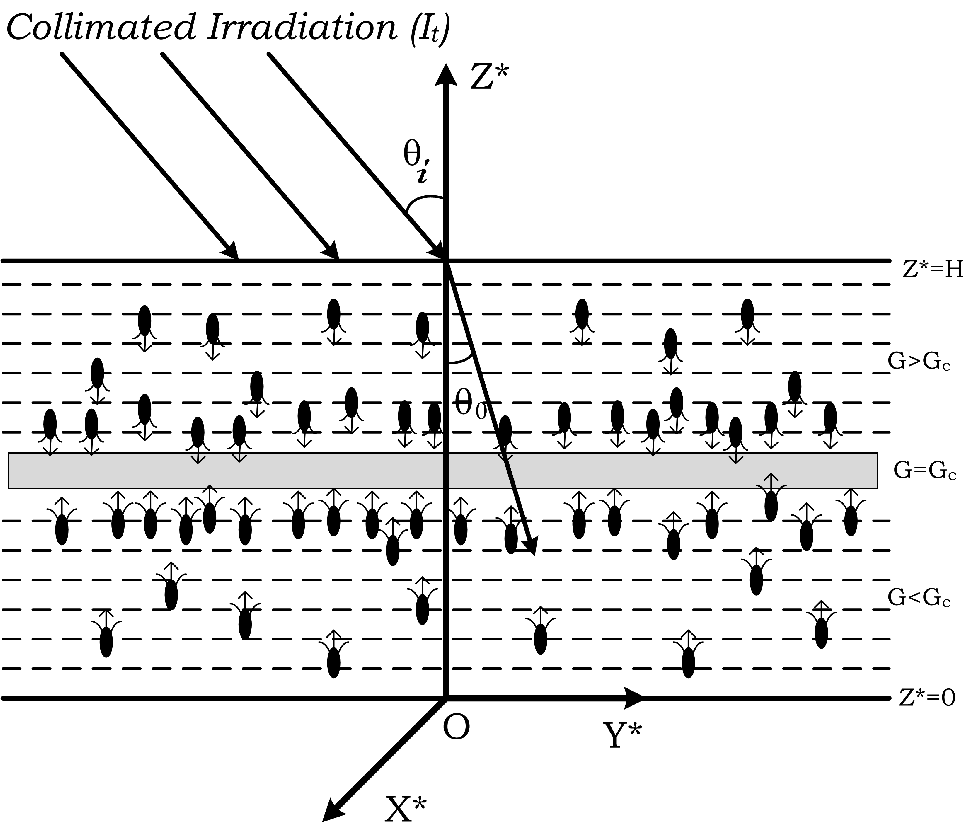}
	\caption{\footnotesize{Formation of the sublayer within the porous medium due to the oblique collimated irradiation at $G=G_c$.}}
	\label{fig1}
\end{figure}

Extensive research has been conducted on the influence of phototaxis on fluid flows within non-porous media. Vincent and Hill~\citep{5vincent1996} investigated the effects of collimated light on bioconvective flow in an algal suspension, followed by Ghorai and Hill~\citep{3ghorai2005}, who analyzed two-dimensional phototactic bioconvection. Both studies neglected the role of light scattering. Later, Ghorai $et$ $al$.\citep{1ghorai2010} and Ghorai and Panda\citep{6ghorai2013} explored the effects of isotropic and anisotropic light scattering, respectively, on bioconvective instability in algal suspensions. Panda and Ghorai~\citep{7panda2013} introduced a two-dimensional model for an isotropically scattering medium, obtaining results that differed from those of Ghorai and Hill~\citep{3ghorai2005}. Additionally, Panda and Singh~\citep{4panda2016} examined bioconvective instability in a non-scattering algal suspension confined between rigid side walls. Panda $et$ $al$.\citep{8panda2016} extended this analysis by studying the impact of diffuse light in an isotropically scattering medium alongside collimated irradiation, while Panda\citep{2panda2020} explored the combined effects of collimated and diffuse irradiation in an anisotropically scattering medium. The influence of oblique light on algal suspensions was later investigated by Panda $et$ $al$.\citep{9panda2022}. The combined effects of diffuse and oblique collimated irradiation on bioconvective instability in isotropically scattering algal suspensions were further examined by Panda and Rajput\citep{17rajput2023}. Subsequently, Rajput and Panda~\citep{18rajput2024} analyzed bioconvective instability under diffuse irradiation alone, excluding collimated irradiation. Further, Rajput and Panda~\citep{38rajput2025b} explored the initiation of fluid motion driven by phototaxis in an isotropically scattering algal suspension with a free top wall, considering only diffuse (scattered) irradiation. Recently, Rajput and Panda~\citep{37rajput2025a} developed a mathematical model to investigate the effects of forward scattering in an algal suspension with a free top wall under diffuse irradiation as the illuminating source. More recently, Rajput and Panda~\citep{39rajput2025c} examined the onset of bioconvective instability in a suspension of forward-scattering phototactic microbes with a rigid top wall exposed exclusively to diffuse irradiation.

Numerous studies have also explored the effects of gravitactic, gyrotactic, and oxytactic microorganisms on fluid dynamics within a porous medium. Kuznetsov and Avramenko~\citep{21kuznetsov2001numerical} investigated gravitactic bioconvection in a porous medium using linear stability analysis to determine the critical permeability. Their findings indicate that bioconvection develops when the permeability of the system exceeds the critical threshold. Kuznetsov and Avramenko~\citep{23kuznetsov2003} extended this research by analyzing bioconvective instability in a porous medium filled with gyrotactic microorganisms. Nield $et$ $al.$\citep{24nield2004} further examined gyrotactic effects in a horizontal porous layer at the onset of bioconvection. Avramenko and Kuznetsov\citep{26avramenko2006} studied the influence of vertical throughflow on the initiation of convection in a suspension of gyrotactic microorganisms within a system of superimposed fluid and porous layers. Similarly, Quang $et$ $al.$\citep{27nguyen2005stability} developed a model to analyze the stability of a gravitactic suspension in a horizontal porous layer, assessing the impact of Rayleigh number and cell swimming speed on system stability. Kumar and Sharma\citep{28kumar2014instability} investigated gyrotactic bioconvection under the combined influence of high-frequency vertical vibration and a porous medium. Thermal effects have also been considered in bioconvection studies within porous media. Kuznetsov~\citep{29kuznetsov2006} analyzed the onset of thermal bioconvection in a shallow fluid-saturated porous layer heated from below, using a suspension of oxytactic microorganisms. Sheremet and Pop~\citep{30sheremet2014thermo} examined thermo-bioconvection in a square porous cavity containing oxytactic microorganisms. Kiran~\citep{31kiran2016throughflow} explored the effects of throughflow and gravity modulation on heat transport in a porous medium. Zhao $et$ $al.$\citep{32zhao2019} developed a model to assess the influence of a porous matrix on thermal bioconvection in a suspension of gyrotactic microorganisms. Biswas $et$ $al.$\citep{33biswas2021thermo} investigated thermo-bioconvection in a porous medium under the combined effects of a magnetic field and oxytactic microorganisms. Recent studies have expanded the understanding of bio-thermal convection in porous systems. Kopp and Yanovsky~\citep{34kopp2023effect} analyzed the impact of gravity modulation on weakly nonlinear bio-thermal convection in a porous medium. They further examined bio-thermal convection in a rotating porous layer saturated with a Newtonian fluid containing gyrotactic microorganisms~\citep{35kopp2023darcy}. Additionally, Kopp and Yanovsky~\citep{36kopp2024weakly} incorporated weakly nonlinear stability analysis to study thermal bioconvection in a porous medium subjected to rotation, gravity modulation, and heat sources.

Phototaxis-driven bioconvection can occur in an isotropic porous medium. In this context, Rajput and Panda~\citep{40rajput2024mathematical} examined the influence of collimated solar irradiation on the onset of phototaxis-driven instability in such a medium. However, under natural conditions, sunlight predominantly strikes surfaces at varying off-normal angles. Despite this, no prior study has explored the onset of phototactic bioconvection in a porous medium while incorporating the effects of oblique collimated irradiation. To address this gap, the present study investigates the impact of oblique collimated irradiation on bioconvection within an isotropic porous medium.

The structure of this article is as follows. First, the proposed phototaxis model is mathematically formulated. Next, the basic state of the governing bioconvective system is established to facilitate the linear stability analysis. The perturbed governing system, derived from the basic state, is then solved numerically. Finally, the numerical results of the proposed model are analyzed and discussed in detail.
%%%%%%%%%%%%%%%%%%%%%%%%%%%%%%%%%%%%%%%%%%%%%%%%%%%%%%%%%%%%%%%%%%%%%%%
\section{Mathematical formulation}
Consider a bioconvective instability driven by phototaxis in a suspension of phototactic microbes. The suspension is confined between two parallel horizontal planes ($z^*=0$ and $z^*=H$) with infinite extent and is subjected to oblique collimated solar irradiation. The light intensity at any specific location $\boldsymbol{x}^*$ within the suspension, in a particular unit direction $\boldsymbol{s}$, is denoted by $I^*(\boldsymbol{x}^*, \boldsymbol{s})$.
%	%%%%%%%%%%%%%%%%%%%%%%%%%%%%%%%%%%%%%%%%%%%%%%%%%%%%%%%%%%%%%%%%%%%%	
\subsection{The governing equations}\label{sec2}
The porous matrix is assumed to have no adverse effects on the microorganisms or their phototactic behavior, including their swimming orientation and speed. For a dilute suspension of swimming microorganisms, the fluid is treated as incompressible. Additionally, the Darcy-Brinkman model is employed in conjunction with the Boussinesq approximation. Under these assumptions, the governing equations for continuity, momentum, and cell conservation are formulated as follows
\begin{equation}\label{1}
\boldsymbol{\nabla}\cdot\boldsymbol{u}^*=0,
\end{equation}
\begin{align}\label{2}
\frac{\rho}{\phi}\left(\frac{\partial\boldsymbol{u}^*}{\partial t^*}+\frac{1}{\phi}\boldsymbol{u}^*\cdot\boldsymbol{\nabla}\boldsymbol{u}^*\right)=-\boldsymbol{\nabla} P_e+\mu{\boldsymbol{\nabla}}^2\boldsymbol{u}^*-\frac{\mu}{K}\boldsymbol{u}^*
-n^*\vartheta \boldsymbol{g}\Delta\rho\hat{\boldsymbol{z}},
\end{align}
\begin{equation}\label{3}
\frac{\partial n^*}{\partial t^*}=-\boldsymbol{\nabla}\cdot\boldsymbol{J}^*,
\end{equation}
where total cell flux is	
\begin{equation*}
\boldsymbol{J}^*=n^*\boldsymbol{u}^*+n^*W_s\boldsymbol{p}-\boldsymbol{D}\boldsymbol{\nabla} n^*.
\end{equation*}

In these equations, $\boldsymbol{u}^*$, $\rho$, and $\mu$ are the velocity, density, and viscosity of the fluid, respectively. Also, $\phi$ represents the porosity of the medium, and $K$ is the permeability of the medium. $P_e$ is the pressure above hydrostatic, $\boldsymbol{g}$ is the gravitational acceleration, $\vartheta$ is the average volume of microorganisms in the suspension, and $n^*$ represents the concentration of microorganisms. Furthermore, the density difference between microorganisms and the base fluid is denoted by $\Delta\rho$, and $\boldsymbol{D}$ represents the diffusivity tensor. Here, it is assumed that $\boldsymbol{D}=DI$ (for more detail see Ref.~\citep{5vincent1996}).  
%	%%%%%%%%%%%%%%%%%%%%%%%%%%%%%%%%%%%%%%%%%%%%%%%%%%%%%%%%%%%%%%%%%%%%%
\subsection{The swimming orientation}	
The radiative transfer equation in the absorbing (non-scattering) medium is given by
\begin{equation}\label{4}
\boldsymbol{s}\cdot\nabla I^*(\boldsymbol{x}^*,\boldsymbol{s})+\kappa I^*(\boldsymbol{x}^*,\boldsymbol{s})=0,
\end{equation}
where $\kappa$ is the extinction coefficient.

%	\begin{figure}[!h]
%		\centering
%		\includegraphics[width=14cm]{oblique_rotation.eps}
%		\caption{\footnotesize{Axial representation of the problem.}}
%		\label{fig1}
%	\end{figure}

At the top layer of the suspension, the light intensity is represented as
\begin{equation*}
I^*(\boldsymbol{x}^*_H,\boldsymbol{s})=\mathrm{I_t}\delta(\boldsymbol{s}-\boldsymbol{s}_0), 
\end{equation*}
where $\boldsymbol{x}^*_H=(x,y,H)$ is a point on the top layer of the suspension. Also, $\mathrm{I_t}$ is the intensity of oblique irradiation.

Consider $\kappa=\alpha n(\boldsymbol{x}^*)$. Thus, Eq.~(\ref{4}) can be rewritten as
\begin{equation}\label{5}
\boldsymbol{s}\cdot\nabla L(\boldsymbol{x}^*,\boldsymbol{s})+\alpha nL(\boldsymbol{x}^*,\boldsymbol{s})=0.
\end{equation}
The value of the total intensity at a given point $\boldsymbol{x}^*$ is determined by 
\begin{equation*}
G(\boldsymbol{x}^*)=\int_0^{4\pi}I^*(\boldsymbol{x}^*,\boldsymbol{s})d\Omega,
\end{equation*}
and similarly, the radiative heat flux is given by 
\begin{equation*}
\boldsymbol{q}(\boldsymbol{x}^*)=\int_0^{4\pi}I^*(\boldsymbol{x}^*,\boldsymbol{s})\boldsymbol{s}d\Omega.
\end{equation*}

The average swimming speed of cells is $W_c$. Thus, the average swimming speed of cells is given by
\begin{equation*}
\boldsymbol{W}_c=W_c\,\boldsymbol{p},
\end{equation*}
where, $\boldsymbol{p}$ is the average swimming orientation, which is determined by
\begin{equation}\label{6}
\boldsymbol{p}=-M(G)\frac{\boldsymbol{q}}{|\boldsymbol{q}|}, 
\end{equation}
where $M(G)$ is the taxis function and it takes the following form 
\begin{equation*}
M(G)=\left\{\begin{array}{ll}\geq 0, & \mbox{ } G(\boldsymbol{x}^*)\leq G_{c}, \\
< 0, & \mbox{ }G(\boldsymbol{x}^*)>G_{c}.  \end{array}\right. 
\end{equation*}

For different types of microbes species, the functional form of the taxis function varies. For example, the typical form of the taxis function is given by 
\begin{equation}\label{7}
M(G)=0.8\sin\left\{\frac{3\pi}{2}\Upsilon(G)\right)-0.1\sin\left(\frac{\pi}{2}\Upsilon(G)\right\},
\end{equation}
where $\Upsilon=Ge^{\beta(G-1)}$ and $\beta$ is related to critical light intensity. 
%%%%%%%%%%%%%%%%%%%%%%%%%%%%%%%%%%%%%%%%%%%%%%%%%%%%%%%%%%%%
\subsection{Boundary conditions}
In our analysis, we shall present results for both free and rigid upper horizontal boundaries. Indeed, in experiments, it is usually the case that the lower boundary is rigid, whilst the upper boundary may be free or rigid.

The boundary conditions for the problem are
\begin{equation}\label{8}
\boldsymbol{u}^*\cdot\hat{\boldsymbol{z}}=0,~~~~ \text{on}~~~~z^*=0,H,
\end{equation}		
\begin{equation}\label{9}
\boldsymbol{J}^*\cdot\hat{\boldsymbol{z}}=0,~~~~ \text{on}~~~~z^*=0,H.
\end{equation}

For a rigid boundary
\begin{equation}\label{10}
\boldsymbol{u}^*\times\hat{\boldsymbol{z}}=0,~~~~ \text{on}~~~~z^*=0,H,
\end{equation}
whilst for a free boundary
\begin{equation}\label{11}
\frac{\partial^2}{\partial z^2}(\boldsymbol{u}^*\cdot\hat{\boldsymbol{z}})=0,~~~~ \text{on}~~~~z^*=0,H.
\end{equation}

The top boundary is exposed to a uniform oblique solar irradiation of magnitude $\mathrm{I_t}$. The top and the bottom boundaries are also
assumed to be non-reflecting, thus,
\begin{equation}\label{12}
I(x^*, y^*, z^*=H, \theta, \phi)=\mathrm{I_t}\delta(\boldsymbol{s}-\boldsymbol{s}_0),~~~  \pi/2\leq\theta\leq\pi,
\end{equation}
\begin{equation}\label{13}
I(x^*, y^*, z^*=0, \theta, \phi) =0,~~~ 0\leq\theta\leq\pi/2. 
\end{equation}

%%%%%%%%%%%%%%%%%%%%%%%%%%%%%%%%%%%%%%%%%%%%%%%%%%%%%%%%%%%%
\subsection{Scaling the equations}
All governing equations are scaled by slandered parameters as used in the phototaxis bioconvection model developed by Rajput and Panda~\citep{37rajput2025a}. Thus, the governing equations become
\begin{equation}\label{14}
\boldsymbol{\nabla}\cdot\boldsymbol{u}=0,
\end{equation}
\begin{equation}\label{15}
\frac{S_c^{-1}}{\phi}\left(\frac{\partial\boldsymbol{u}}{\partial t}+\frac{1}{\phi}\boldsymbol{u}\cdot\boldsymbol{\nabla}\boldsymbol{u}\right)=-\boldsymbol{\nabla} P_{e}+\boldsymbol{\nabla}^{2}\boldsymbol{u}-D_a^{-1}\boldsymbol{u}+Rn\hat{\boldsymbol{z}},
\end{equation}
\begin{equation}\label{16}
\frac{\partial n}{\partial t}=-\boldsymbol{\nabla}\cdot(n   \boldsymbol{u}+nV_{c}\boldsymbol{p}-\boldsymbol{\nabla}n),
\end{equation}
where $S_c=\nu/D$ is the Schmidt number, $V_c=W_sH/D$ is dimensionless swimming speed, $R=\bar{n}\vartheta g\Delta{\rho}H^{3}/\nu\rho{D}$ is the bio-convective Rayleigh number, and $D_a=K/H^2$ is the the Darcy number.

After scaling, the boundary conditions become
\begin{equation}\label{17}
\boldsymbol{u}\cdot\hat{\boldsymbol{z}}=0,~~~~ \text{on}~~~~z=0,1,
\end{equation}		
\begin{equation}\label{18}
\boldsymbol{J}\cdot\hat{\boldsymbol{z}}=0,~~~~ \text{on}~~~~z=0,1.
\end{equation}

For a rigid boundary
\begin{equation}\label{19}
\boldsymbol{u}\times\hat{\boldsymbol{z}}=0,~~~~ \text{on}~~~~z=0,1,
\end{equation}
whilst for a free boundary
\begin{equation}\label{20}
\frac{\partial^2}{\partial z^2}(\boldsymbol{u}\cdot\hat{\boldsymbol{z}})=0,~~~~ \text{on}~~~~z=0,1.
\end{equation}

Next, the RTE becomes after scaling
\begin{equation}\label{21}
\boldsymbol{s}\cdot\nabla I(\boldsymbol{x},\boldsymbol{s})+\tau_H nL(\boldsymbol{x},\boldsymbol{s})=0.
\end{equation}
where $\tau_H=\alpha\tilde{n}H$ is the scaled extinction coefficient. After scaling the light intensity at the top and bottom becomes
\begin{equation}\label{22}
I(x, y, z=1, \theta, \phi)=\mathrm{I_t}\delta(\boldsymbol{s}-\boldsymbol{s}_0),~~~  \pi/2\leq\theta\leq\pi,
\end{equation}
\begin{equation}\label{23}
I(x, y, z=0, \theta, \phi) =0,~~~ 0\leq\theta\leq\pi/2. 
\end{equation}

Note that an unstarred symbol is used here for the scaled variables.
%%%%%%%%%%%%%%%%%%%%%%%%%%%%%%%%%%%%%%%%%%%%%%%%%%%%%%%%%%%%%%%%%%%%%%%%%%%	
\section{THE BASIC SOLUTION}	
Equations (\ref{14})--(\ref{16}) and (\ref{21}) together with the boundary conditions possess a static equilibrium solution in which	
\begin{equation*}
\boldsymbol{u}=0,\,n=n_s(z),\,I=I_s(z,\theta)~~ \text{and}~~ \boldsymbol{p}=\boldsymbol{p}_s.
\end{equation*}

The total intensity at the basic state is given by
\begin{equation*}
G_s(z)=\int_0^{4\pi}I_s(z,\theta)d\Omega.
\end{equation*}

If horizontal homogeneity and azimuthally isotropic are assumed
(as considered here), the RTE becomes
\begin{equation}\label{24}
\frac{\partial I_s(z,\theta)}{\partial z}+\frac{\tau_H n_sI_s(z,\theta)}{\cos\theta}=0,
\end{equation}	
with the top boundary condition 
\begin{equation}\label{25}
I_s(z= 1, \theta) =\mathrm{I_t}\delta(\boldsymbol{s}-\boldsymbol{s}_0).
\end{equation}

After calculations, we get 
\begin{equation*}
I_s(z,\theta)=\mathrm{I_t}\exp\left(\int_z^1\frac{\tau_H n_s(z')}{\cos\theta}dz'\right)\delta(\boldsymbol{s}-\boldsymbol{s}_0).
\end{equation*}

The total intensity in the equilibrium state can be written as
\begin{equation*}
G_s(z)=\int_0^{4\pi}I_s(z,\theta)d\Omega=\mathrm{I_t}\exp\left(-\int_z^1\frac{\tau_H n_s(z')dz'}{\cos\theta_0}\right).
\end{equation*}

Also, the radiative heat flux in the basic state can be determined by	
\begin{equation*}
\boldsymbol{q}_s(z)=\int_0^{4\pi}I_s(z,\theta)\boldsymbol{s}d\Omega=\left\{-\mathrm{I_t}\exp\left(-\int_z^1\frac{\tau_H n_s(z')dz'}{\cos\theta_0}\right)\cos\theta_0\right\}\hat{\boldsymbol{z}}.
\end{equation*}

Since the illuminating source lies in the opposite direction to the
radiative heat flux vector, the mean swimming direction in the basic
state is expressed as
\begin{equation*}
\boldsymbol{p_s}=M_s\hat{\boldsymbol{z}},
\end{equation*}	
where $M_s=M(G_s).$

The basic concentration satisfies	
\begin{equation}\label{26}
\frac{dn_s}{dz}-V_cM_sn_s=0,
\end{equation}
which is supplemented by the cell conservation relation
\begin{equation}\label{27}
\int_0^1n_s(z)dz=1.
\end{equation}

Now introducing a new variable $\varpi=-\int_z^1 n_s dz$, Eq.~(\ref{26}) becomes	
\begin{equation}\label{28}
\frac{d^2\varpi}{dz^2}-V_cM_s\frac{d\varpi}{dz}=0,
\end{equation}
subject to the boundary conditions
\begin{equation*}
\varpi+1=0~~~~~~~~~~~~~\text{on}~~~~~z=0,
\end{equation*}
\begin{equation}\label{29}
\varpi=0~~~~~~~~~~~~~~~~~~~~\text{on}~~~~~z=1.
\end{equation}

Thus, the total intensity in terms of $\varpi$ can be written as
\begin{equation*}
G_s(z)=\mathrm{I_t}\exp{\left(\frac{\tau_H \varpi}{\cos\theta_0}\right)}.
\end{equation*}

\begin{figure}[!htbp]
	\centering
	\includegraphics[scale=0.53]{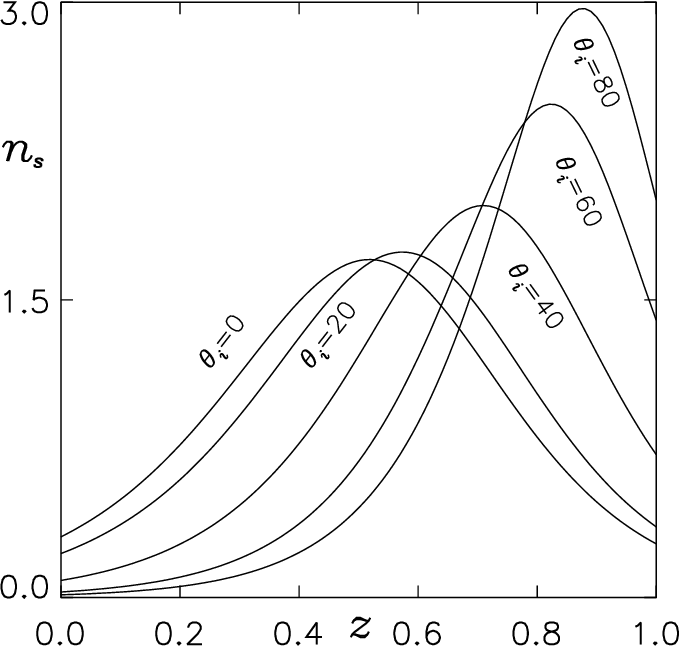}
	\caption{\footnotesize{Effects of the angle of incidence $\theta_i$ (deg) on the base concentration profiles. Here the parameters $V_c=10$ , $\tau_H=0.5$, $\mathrm{I_t}=0.8$, and $G_c=0.63$ are kept fixed.}}
	\label{fig2}
\end{figure}

Equations (\ref{28}) and (\ref{29}) form a boundary value problem, and the solution to this problem is obtained using a numerical technique called the shooting method.

Figure \ref{fig2} shows the basic concentration profiles with the effects of angle of incidence for $V_c=10$, $\tau_H=0.5$, $\mathrm{I_t}=0.8$, and $G_c=0.63$. As the angle of incidence increases, the location of maximum concentration shifts towards the top of the suspension. Additionally, the profiles become more steep as the angle of incidence increases.

%%%%%%%%%%%%%%%%%%%%%%%%%%%%%%%%%%%%%%%%%%%%%%%%%%%%%%%%%%%%%%%%%%%%%%%%%%	
\section{Linear stability of the problem}
To perform linear stability analysis, the basic state is perturbed
via infinitesimal disturbances as
\begin{align}\label{30}
\nonumber\boldsymbol{u}=0+{\epsilon} \boldsymbol{u}_1+\mathcal{O}({\epsilon}^2),~n=n_s+{\epsilon} n_1+\mathcal{O}({\epsilon}^2),\\I=I_s+{\epsilon} I_1+\mathcal{O}({\epsilon}^2),~\boldsymbol{p}=\boldsymbol{p}_s+ {\epsilon} {\boldsymbol{p}}_1 +\mathcal{O}({\epsilon}^2),
\end{align}
where $\boldsymbol{u}_1=(u_1,v_1,w_1)$.

The linearized equations about the basic state are
\begin{equation}\label{31}
\boldsymbol{\nabla}\cdot \boldsymbol{u}_1=0,
\end{equation}
\begin{equation}\label{32}
\frac{S_c^{-1}}{\phi}\left(\frac{\partial \boldsymbol{u_1}}{\partial t}\right)=-\boldsymbol{\nabla} P_{e}+\boldsymbol{\nabla}^{2}\boldsymbol{u}_1-D_a^{-1}\boldsymbol{u}_1-Rn_1\hat{\boldsymbol{z}},
\end{equation}
\begin{equation}\label{33}
\frac{\partial{n_1}}{\partial{t}}+V_c\boldsymbol{\nabla}\cdot(\boldsymbol{p}_sn_1+\boldsymbol{p}_1n_s)+w_1\frac{dn_s}{dz}=\boldsymbol{\nabla}^2n_1.
\end{equation}

If $G=G_s+{\epsilon}G_1+\mathcal{O}({\epsilon}^2)$, then the total intensity in the basic state is perturbed as $\mathrm{I_t}\exp\left(\frac{-\tau_H\int_z^1(n_s(z')+{\epsilon} n_1+\mathcal{O}({\epsilon}^2))dz'}{\cos\theta_0}\right)$ and after simplification, we get
\begin{equation}\label{34}
G_1=\mathrm{I_t}\exp\left(\frac{\int_1^z \tau_H n_s(z')dz'}{\cos\theta_0}\right)\left(\frac{\int_1^z\tau_H n_1 dz'}{\cos\theta_0}\right),
\end{equation}

The perturbed mean swimming orientation for a non-scattering algal suspension is expressed as
\begin{equation}\label{35}
\boldsymbol{p}_1=G_1\frac{dM_s}{dG}\hat{\boldsymbol{z}}.
\end{equation}

We eliminate $P_e$ and the horizontal component of $\boldsymbol{u}_1$by taking
the curl of Eq.~(\ref{32}) twice and retaining the $z$-component of the result.
Then, Eqs.~(\ref{31})--(\ref{33}) reduce to two equations for $w_1$ and $n_1$.These variables can be decomposed into normal modes as
\begin{equation*}
w_1=W(z)\exp{(\sigma t+i(lx+my))}, 
\end{equation*}
\begin{equation*}
n_1=\Theta(z)\exp{(\sigma t+i(lx+my))}.  
\end{equation*}

The linear stability equations become
\begin{equation}\label{36}
\left(\sigma S_c^{-1}+k^2-D^2\right)\left( D^2-k^2\right)W(z)=Rk^2\Theta(z),
\end{equation}
\begin{align}\label{37}
\nonumber\aleph_1(z)\int_z^1\Theta(z') dz'+(\sigma+k^2+\aleph_2(z))\Theta(z)\\+\aleph_3(z)D\Theta(z)-D^2\Theta(z)=-Dn_sW(z), 
\end{align}
where
\begin{subequations}\label{38}
	\begin{equation}
	\aleph_1(z)=-(\tau_H/\cos\theta_0) V_cD\left(n_sG_s\frac{dM_s}{dG}\right),
	\end{equation}
	\begin{equation}
	\aleph_2(z)=2(\tau_H/\cos\theta_0) V_c n_s G_s\frac{dM_s}{dG},
	\end{equation}
	\begin{equation}
	\aleph_3(z)=V_cM_s,
	\end{equation}
\end{subequations}
subject to the boundary conditions
\begin{align}\label{39}
\nonumber W=\frac{dW}{dz}=\frac{d\Theta}{dz}-\aleph_3(z)\Theta+n_sV_c(\tau_H/\cos\theta_0)\\
\times G_s\left(\int_z^1\Theta(z')dz'\right)\frac{dM_s}{dG}=0, ~~~\text{on}~~~z=0,
\end{align}
\begin{equation}\label{40}
W=\frac{d^2W}{dz^2}=\frac{d\Theta}{dz}-\aleph_3(z)\Theta=0~~~\text{on}~~~z=1.
\end{equation}

At a rigid upper surface, the condition in Eq.~(\ref{40}) is replaced by
\begin{equation}\label{41}
W=\frac{dW}{dz}=\frac{d\Theta}{dz}-\aleph_3(z)\Theta=0~~~\text{on}~~~z=1.
\end{equation}

Here, $k=\sqrt{(l^2+m^2)}$ is the horizontal wavenumber and it represents the modulation of bioconvection pattern in horizontal directions.
Equations (\ref{36})--(\ref{37}) form an eigenvalue problem for $\sigma$ as a function of the dimensionless parameters $\theta_0,S_c,V_c,k,\tau_H,D_a$ and $R$. The basic state becomes unstable whenever Re$\sigma> 0$.

Now, introducing a new variable as
\begin{equation}\label{42}
\Phi(z)=\int_1^z\Theta(z')dz',
\end{equation}

The perturbed system becomes
\begin{align}\label{43}
\left(\sigma \frac{S_c^{-1}}{\phi}+D_a^{-1}+k^2-D^2\right)\left( D^2-k^2\right)W=Rk^2D\Phi,
\end{align}
\begin{align}\label{44}
\nonumber\aleph_1(z)\Phi+(\sigma+k^2+\aleph_2(z))D\Phi+\aleph_3(z)D^2\Phi-D^3\Phi\\=-Dn_sW, 
\end{align} 
with the boundary conditions	
\begin{align}\label{45}
{W}=DW=\aleph_2(z)\Phi+2\aleph_3(z)D\Phi-2D^2\Phi=0,~~~\text{on}~~z=0,
\end{align}	
\begin{align}\label{46}
{W}=D^2W=\aleph_2(z)\Phi+2\aleph_3(z)D\Phi-2D^2\Phi=0,~~\text{on}~~z=1,
\end{align}	
and for a rigid upper surface, the condition in Eq.~(\ref{46}) is replaced by
\begin{align}\label{47}
{W}=DW=\aleph_2(z)\Phi+2\aleph_3(z)D\Phi-2D^2\Phi=0,~~\text{on}~~z=1.
\end{align}	

There is also an extra boundary condition
\begin{equation}\label{48}
\Phi(z)=0~~~~~~~~\text{on}~~~z=1,
\end{equation}	
which follows from Eq.~(\ref{42}). The new aspects of the proposed model
via the effects of oblique collimated irradiation are incorporated in the
terms $\aleph_1, \aleph_2$ and $\aleph_3$ which are found on the perturbed cell conservation equation, Eq.~(\ref{44}). Kindly note that we can revert back to the phototaxis model proposed by Rajput and Panda~\citep{40rajput2024mathematical} as the angle of incidence $\theta_i$ (deg) of the oblique collimated irradiation tends to zero.
%%%%%%%%%%%%%%%%%%%%%%%%%%%%%%%%%%%%%%%%%%%%%%%%%%%%%%%%%%%%%%%%%%%%%%%%%%%	
\section{Solution technique and numerical results}\label{sec5}
A fourth-order accurate finite difference scheme, based on the Newton-Raphson-Kantorovich (NRK) iterations~\citep{11cash1980}, is employed to solve the system of Eqs.~(\ref{43})--(\ref{44}). This scheme facilitates the computation of neutral stability curves in the $(k, R)$--plane for various parameter sets. The neutral curve $R^{(n)}(k)$ consists of multiple branches, each corresponding to a potential solution to the linear stability problem. The most critical branch, identified by the lowest value of $R$, represents the most unstable bioconvective solution, denoted as $(k_c, R_c)$, where $R_c$ is the critical bioconvective Rayleigh number and $k_c$ is the corresponding wavenumber.

To systematically explore the complex parameter space, we analyze the impact of the Darcy number on bioconvective instability while keeping certain parameters constant and varying others. This methodology enables a focused examination of specific system components and their individual contributions to the onset of bioconvection in a porous medium. For model reliability, the Schmidt number is fixed at $S_c=20$, the light source intensity at $\mathrm{I_t}=0.8$, and the porosity of the medium at $\phi=0.76$. The extinction coefficient is set to $\tau_H=0.5$, the cell swimming speed to $V_c=10$, and the angle of incidence $\theta_i$ varies from 0 to 80 degrees. Additionally, the critical light intensity is chosen such that, at $\theta_i=0$, the sublayer in the basic state forms at the mid-height of the suspension.		
%%%%%%%%%%%%%%%%%%%%%%%%%%%%%%%%%%%%%%%%%%%%%%%%%%%%%%%%%%%%%%%%%%%%%%%%%%%%	
\subsection{$V_c=15$}
$(a)$ \textit{When top surface is stress-free:}\\

\begin{figure}[!htbp]
	\centering
	\includegraphics[scale=0.5]{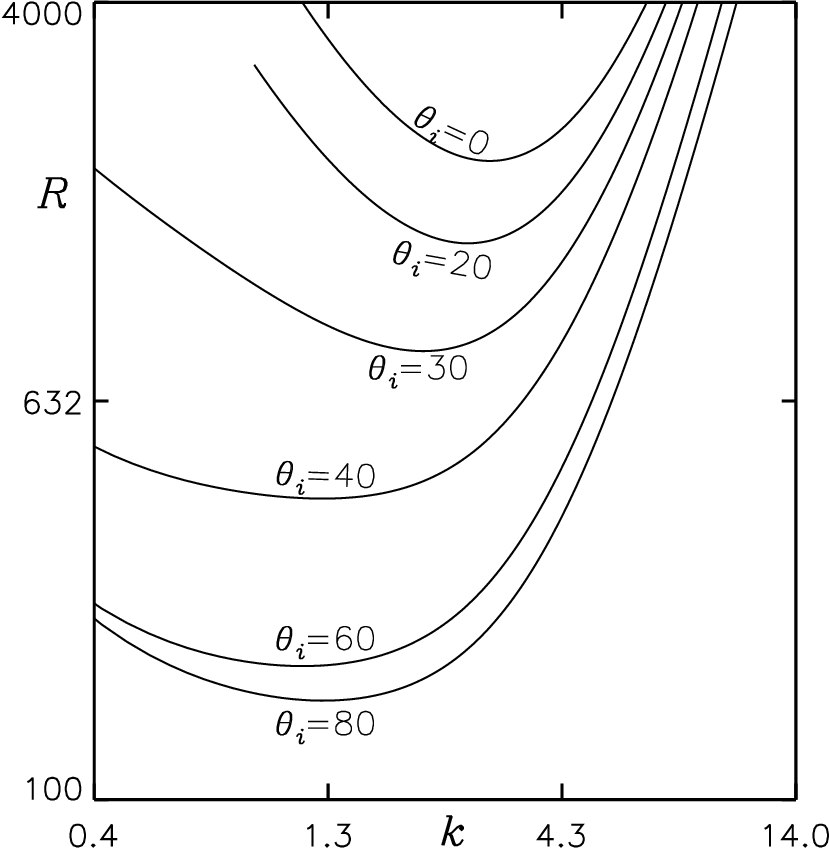}
	\caption{\label{fig3} Effects of the angle of incidence on the neutral curves for the fixed parameters $V_c=10$, $\tau_H=0.5$, $\mathrm{I_t}=0.8$, $D_a=0.1$, and $G_c=0.63$. Here, the top surface is stress-free.}
\end{figure}

Fig.~\ref{fig3} shows the neutral stability curves for different values of angle of incidence when the top surface of the suspension is stress-free. Here the other parameters $V_c=10$ , $\tau_H=0.5$, $\mathrm{I_t}=0.8$, $D_a=0.1$, and $G_c=0.63$ are kept fixed. The critical Rayleigh number decreases as the angle of incidence increases. Thus, the suspension becomes more unstable as the angle of incidence increases.

\begin{figure}[!htbp]
	\centering
	\includegraphics[scale=0.5]{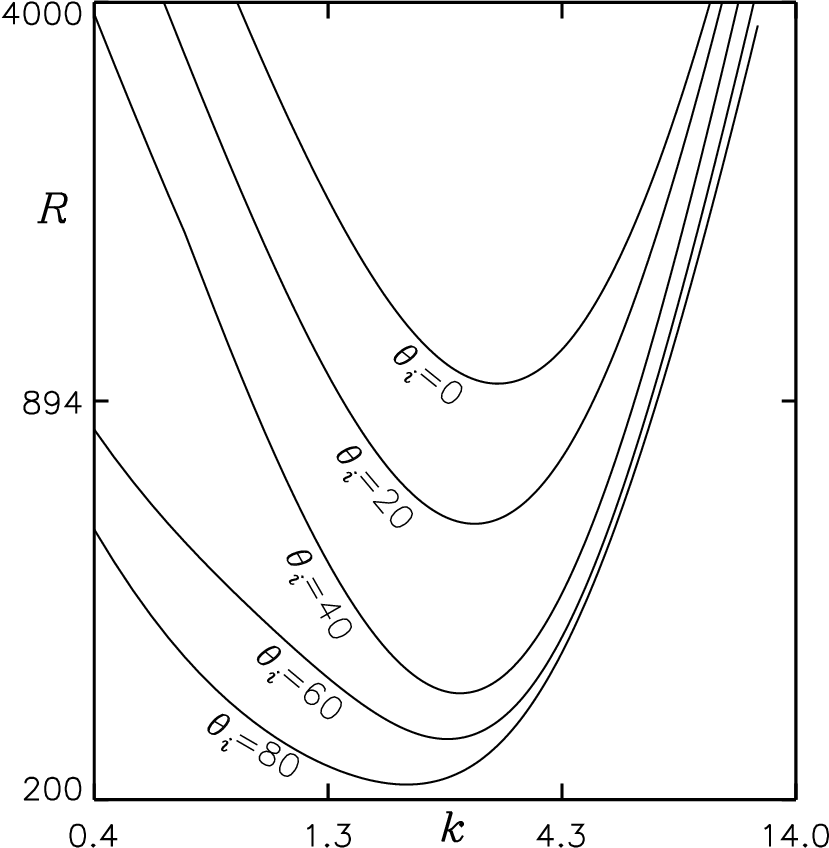}
	\caption{\label{fig4}Effects of the angle of incidence on the neutral curves for fixed parameters $V_c=10$, $\tau_H=0.5$, $\mathrm{I_t}=0.8$, $D_a=0.1$, and $G_c=0.63$. Here, the top surface is rigid.}
\end{figure}

$(b)$ \textit{When top surface is rigid:}\\
Fig.~\ref{fig4} shows the neutral stability curves for different values of angle of incidence when the top surface of the suspension is rigid. Here the other parameters $V_c=10$ , $\tau_H=0.5$, $\mathrm{I_t}=0.8$, $D_a=0.1$, and $G_c=0.63$ are kept fixed. The critical Rayleigh number decreases as the angle of incidence increases. Thus, the suspension becomes more unstable as the angle of incidence increases.

%%%%%%%%%%%%%%%%%%%%%%%%%%%%%%%%%%%%%%%%%%%%%%%%%%%%%%%%%%%%%%%%%%%%%%%%%	
\section{Conclusion}\label{sec7}

In this study, we propose a novel model for light-induced bioconvection within an isotropic porous medium, incorporating the effects of oblique collimated irradiation for the first time. This model considers the influence of the angle of incidence on suspension behavior, where light enters obliquely from the top. The governing equations are formulated using the Darcy-Brinkman model. The results derived from the linear stability analysis of this system are summarized as follows.

In the basic state, it is observed that self-shading becomes more pronounced with an increasing angle of incidence due to an extended slant-path length. As a result, algae at a fixed interior depth receive lower light intensities, leading to a shift in the location of maximum basic concentration toward the suspension top. Additionally, the magnitude of the maximum basic concentration increases as the angle of incidence rises.

The linear stability analysis reveals that the critical Rayleigh number decreases with an increasing angle of incidence for a fixed Darcy number, indicating that the system becomes more unstable under such conditions. Furthermore, it is observed that the critical Rayleigh number is higher in the case of a rigid top wall compared to a stress-free top wall. This suggests that the suspension exhibits greater stability when a rigid top boundary is present. 	
%%%%%%%%%%%%%%%%%%%%%%%%%%%%%%%%%%%%%%%%%%%%%%%%%%%
\section*{Author declarations}
\section*{Conflict of interest}
The authors have no conflicts to disclose.
\section*{Data availability}
The data that support the findings of this study are available within the article.
%\section*{BIBLIOGRAPHY}
\bibliography{POROUS_OBLIQUE}	
	
\end{document}